\documentclass[aps,prl,twocolumn,groupedaddress,showpacs]{revtex4}

\usepackage{bm}
\usepackage{graphicx}

\begin{document}

\title{Unconventional Anisotropic $s$-Wave Superconducting Gaps of LiFeAs Iron-Pnictide Superconductor}

\author{
	K. Umezawa,$^{1}$
	Y. Li,$^{2}$
	H. Miao,$^{3}$
	K. Nakayama,$^{1}$
	Z.-H. Liu,$^{2}$
	P. Richard,$^{3}$
	T. Sato,$^{1,4}$
	J. B. He,$^{2}$
	D.-M. Wang,$^{2}$
	G. F. Chen,$^{2}$
	H. Ding,$^{3}$
	T. Takahashi,$^{1,5}$
	and S.-C. Wang$^{2}$}

\affiliation{$^1$Department of Physics, Tohoku University, Sendai 980-8578, Japan}
\affiliation{$^2$Department of Physics, Renmin University of China, Beijing 100872, China}
\affiliation{$^3$Beijing National Laboratory for Condensed Matter Physics, and Institute of Physics, Chinese Academy of Sciences, Beijing 100190, China}
\affiliation{$^4$TRiP, Japan Science and Technology Agency (JST), Kawaguchi 332-0012, Japan}
\affiliation{$^5$WPI Research Center, Advanced Institute for Materials Research, Tohoku University, Sendai 980-8577, Japan}

\date{\today}

%\begin{minipage}[t]{6.8in}
\begin{abstract}
We have performed high-resolution angle-resolved photoemission spectroscopy on Fe-based superconductor LiFeAs ($T_{\rm c}$ = 18 K). We reveal multiple nodeless superconducting (SC) gaps with 2$\Delta$/$k_{\rm B}$$T_{\rm c}$ ratios varying from 2.8 to 6.4, depending on the Fermi surface (FS). We also succeeded in directly observing a gap anisotropy along the FS with magnitude up to $\sim$30 \%. The anisotropy is four-fold symmetric with an antiphase between the hole and electron FSs, suggesting complex anisotropic interactions for the SC pairing. The observed momentum dependence of the SC gap offers an excellent opportunity to investigate the underlying pairing mechanism.
\end{abstract}


%\end{minipage}
\maketitle
%\narrowtext

The discovery of Fe-based superconductors \cite{Kamihara} generated intensive debates on the superconducting (SC) mechanism. The SC gap, which characterizes the energy cost for breaking a Cooper pair, is an important quantity to clarify the SC mechanism. The gap size and its momentum dependence reflect the strength and anisotropy of the pairing interactions, respectively. Although conventional phonon-mediated superconductors exhibit a $s$-wave SC gap with a 2$\Delta$/$k_{\rm B}$$T_{\rm c}$ ratio close to 3.5, no consensus has been reached on the SC gap character in the newly discovered Fe-based superconductors. Motivated by high-$T_{\rm c}$ values up to 56 K \cite{56K}, the possibility of unconventional superconductivity has been intensively discussed. A plausible candidate is the SC pairing mediated by antiferromagnetic (AF) interactions.  Two different approaches, based on the itinerant spin fluctuations promoted by Fermi-surface (FS) nesting \cite{Mazin, Kuroki} and the local AF exchange couplings \cite{Seo}, predict the so-called $s_{\pm}$-wave pairing state, in which the gap shows a $s$-wave symmetry that changes sign between different FSs. Owing to the multi-orbital nature and the characteristic crystal symmetry of Fe-based superconductors, $s$-wave pairing originating from novel orbital fluctuations has been also proposed \cite{Kontani, Yanagi}. In addition, ferromagnetic interactions may lead to $p$-wave superconductivity if the electronic structure satisfies a specific condition \cite{FM}. The unconventional nature of the superconductivity is supported by experimental observations such as the strongly FS dependent anomalously large SC gaps \cite{Hong, Zhou, Hasan, Nakayama1, Terashima, Nakayama2, Miao, Li} and the possible sign change in the gap function \cite{Resonance, SQUID, STM} on moderately doped BaFe$_2$As$_2$, NdFeAsO and FeTe$_{1-x}$Se$_x$.  However, recent experimental reports on LiFeAs indicated nearly isotropic $s$-wave gap with much smaller 2$\Delta$/$k_{\rm B}$$T_{\rm c}$ value of $\sim$3.5 \cite{Inosov, Borisenko1}. These results seem rather consistent with conventional superconductivity, thus questioning whether the SC mechanism in Fe-based superconductors is conventional and universal. To get an insight into the SC mechanism of Fe-based superconductors, further experimental investigations of the SC gap on LiFeAs are indispensable.

In this Letter, we report the detailed SC gap character of LiFeAs ($T_{\rm c}$ = 18 K) studied by high-resolution angle-resolved photoemission spectroscopy (ARPES), which is a unique technique to directly observe the momentum ($k$) resolved SC gap. We find the opening of larger (smaller) SC gap on smaller (larger) FS, in agreement with the gap function derived from the AF interactions. Moreover, we demonstrate experimental evidence for strong-coupling behavior and a moderate gap anisotropy along some of the FS sheets. These results unambiguously indicate the unconventional nature of the superconductivity in LiFeAs.

High-quality single crystals of LiFeAs ($T_{\rm c}$ = 18 K) were grown by the self-flux method \cite{Sample}. Ultrahigh-resolution ARPES measurements were performed at Tohoku University using a VG-SCIENTA SES2002 spectrometer with a high-flux He discharge lamp ($h$$\nu$ = 21.218 eV). The energy resolution was set at 1.5 meV and 12 meV for SC gap measurements and for band and FS mapping, respectively, and the angular resolution was set at 0.2$^{\circ}$. Fresh surfaces for the ARPES measurements were obtained by cleaving crystals $in$ $situ$ in a working vacuum better than 4$\times$10$^{-11}$ Torr. The Fermi energy ($E_{\rm F}$) of the samples was referenced to that of a gold film evaporated onto the sample holder.

\begin{figure*}[!t]
\begin{center}
\includegraphics[width=6in]{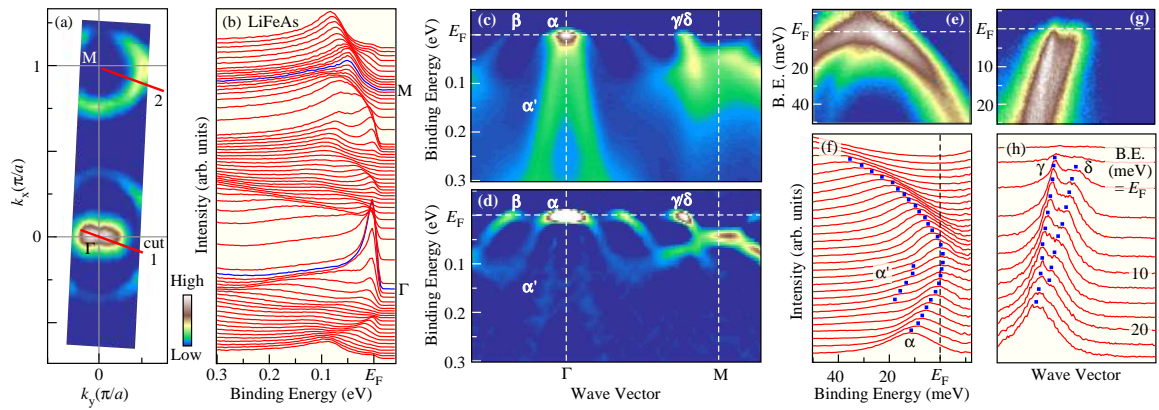}
\end{center}
\caption{
(Color online) (a) Plot of the ARPES intensity at $E_{\rm F}$ of LiFeAs ($T_{\rm c}$ = 18 K) as a function of the two-dimensional wave vector measured with the He I$\alpha$ line ($h$$\nu$ = 21.218 eV). The intensity is obtained by integrating the spectra within $\pm$5 meV with respect to $E_{\rm F}$. (b) ARPES spectra along the $\Gamma$-M high-symmetry line. (c) and (d) Intensity plot and second-derivative intensity plot of (b), respectively, as a function of binding energy and wave vector. (e) ARPES intensity plot at $T$ = 50 K divided by a Fermi-Dirac function measured along cut 1 in (a), and (f) corresponding energy distribution curves. (g) ARPES intensity plot at 20 K along cut 2 and (h) corresponding momentum distribution curves. Blue dots in (f) and (h) are guides for the eye to trace the band dispersion.
}
\end{figure*}

Figure 1(a) displays the ARPES intensity at $E_{\rm F}$ of LiFeAs plotted as a function of the two-dimensional wave vector. We find a bright intensity spot at the $\Gamma$ point in addition to the relatively large FSs centered at the $\Gamma$ and M points. The band dispersion along the $\Gamma$-M high-symmetry line in Figs. 1(b)-1(d) shows that there are three holelike bands centered at the $\Gamma$ point, the outermost $\beta$ band forming the large FS visible in Fig. 1(a). The band maxima of the other two bands ($\alpha$ and $\alpha$' bands) are located very close to $E_{\rm F}$, producing the bright spot in Fig. 1(a). To clarify whether the $\alpha$ and $\alpha$' bands are touching $E_{\rm F}$ or not, we have carefully traced their dispersions by dividing ARPES spectra at 50 K by the Fermi-Dirac distribution function [Figs. 1(e) and 1(f)]. The results show that the $\alpha$ band produces a small FS, whereas the $\alpha$' band sinks below $E_{\rm F}$ by $\sim$10 meV. At the M point, we observed two electron pockets (called $\gamma$ and $\delta$), as demonstrated in Figs. 1(g) and 1(h).  These results indicate that there are two holelike and two electronlike FSs centered at the $\Gamma$ and M points, respectively, which is consistent with a previous ARPES study \cite{Borisenko1}. The hole and electron carrier numbers estimated from the FS volume (0.2 holes/Fe and 0.18 electrons/Fe, respectively) are nearly compensated, suggesting the non-carrier-doped intrinsic nature of the LiFeAs sample.

\begin{figure}[!t]
\begin{center}
\includegraphics[width=3in]{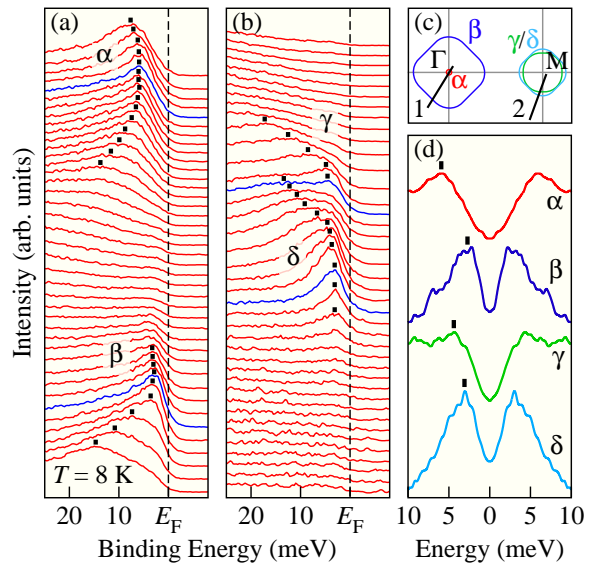}
\end{center}
\caption{
(Color online) (a) and (b) High-resolution ARPES spectra in the SC state (8 K) measured along cut 1 and 2 in (c), respectively. The ARPES spectra at $k_{\rm F}$ points are indicated by blue curves. Dots are guides for the eye to trace the band dispersion. (c) Schematic FS and $k$ location of the cuts. (d) Symmetrized ARPES spectra in the SC state measured at $k_{\rm F}$ points of the $\alpha$, $\beta$, $\gamma$ and $\delta$ bands.
}
\end{figure}

To elucidate the SC gap character of LiFeAs, we have performed ultrahigh-resolution ARPES measurements near $E_{\rm F}$ in the SC state. Figure 2(a) shows the ARPES spectra recorded near the $\Gamma$ point at 8 K. In contrast to the data in the normal state, both the $\alpha$ and $\beta$ bands exhibit a gap opening evidenced by a shift of the leading-edge midpoint toward higher binding energy ($E_{\rm B}$). The leading-edge shift of the $\alpha$ band (about 2.2 meV) is larger than that for the $\beta$ band (0.6 meV), suggesting the FS dependence of the SC gap. We also observed a signature of the FS-dependent SC gap on the electronlike FSs [Fig. 2(b)], although the difference of gap size is smaller than that for the holelike FSs. To highlight the FS-dependent SC gap among four FSs, we directly compare ARPES spectra measured at Fermi wave-vector ($k_{\rm F}$) points in Fig. 2(d). Each spectrum has been symmetrized with respect to $E_{\rm F}$ to eliminate the effect of the Fermi-Dirac distribution function. All the spectra clearly show two-peaked structure, indicative of the gap opening. The SC gap values ($\Delta$) obtained by numerical fitting with the BCS spectral function \cite{Norman} are 5.0, 2.5, 4.2 and 2.8 meV for the $\alpha$, $\beta$, $\gamma$ and $\delta$ bands, respectively (note that the gap value is larger on smaller FS). The corresponding 2$\Delta$/$k_{\rm B}$$T_{\rm c}$ ratios are 6.4, 3.2, 5.4 and 3.6, demonstrating strong-coupling superconductivity in LiFeAs.  While the previous study, which defined the gap size using the leading-edge shift, suggested a weak-coupling behavior in LiFeAs \cite{Borisenko1}, we caution that the leading-edge gap underestimates the SC gap size and the true gap size should be estimated by numerical fitting using the BCS spectral function. Thus the observed anomalously large 2$\Delta$/$k_{\rm B}$$T_{\rm c}$ ratio exceeding 6 is likely an essential property of LiFeAs.

\begin{figure}[!t]
\begin{center}
\includegraphics[width=3in]{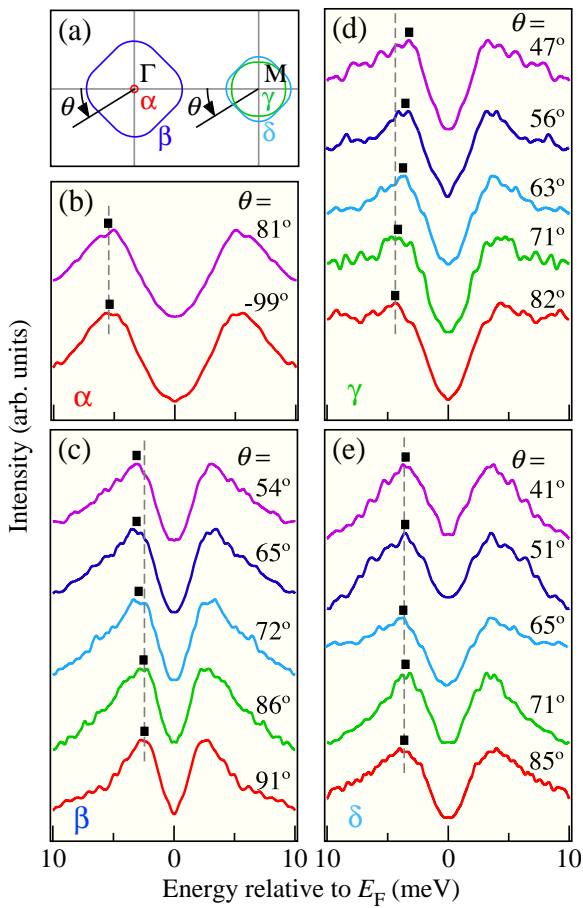}
\end{center}
\caption{
(Color online) (a) Schematic FS and definition of the FS angle ($\theta$). (b)-(e) Symmetrized ARPES spectra in the SC state measured at various $k_{\rm F}$ points of the $\alpha$, $\beta$, $\gamma$ and $\delta$ bands. Dashed lines and dots are guides for the eye.
}
\end{figure}

\begin{figure}[!t]
\begin{center}
\includegraphics[width=3in]{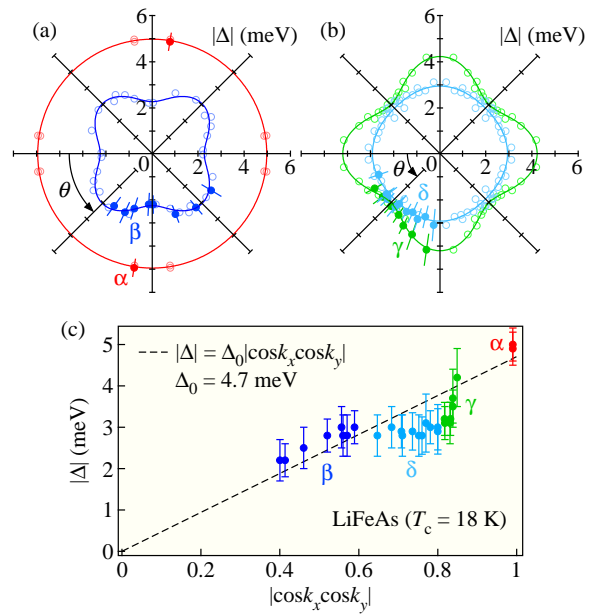}
\end{center}
\caption{
(Color online) (a) and (b) Polar plots of the SC gap size for the $\alpha$/$\beta$ and $\gamma$/$\delta$ FSs, respectively, as a function of $\theta$ defined in Fig. 3(a). Filled circles are the original data, and open circles are the folded data which take into account the four-fold symmetry. Error bars in (a)-(c) originate from the fitting uncertainty on $\Delta$, as well as experimental uncertainties in determining the energy position of $E_{\rm F}$ and the momentum location of $k_{\rm F}$ (less than 0.1 and 0.2 meV, respectively). Solid curves show the fitting results with $\Delta$($\theta$) = $\Delta_0$+$\Delta_1$cos[4($\theta$+$\phi$)]. (c) Plot of the SC gap size as a function of $\mid$cos$k_x$cos$k_y$$\mid$. The fitting result assuming the gap function $\mid$$\Delta$$\mid$ = $\Delta_0$$\mid$cos$k_x$cos$k_y$$\mid$ is indicated by a black dashed line.
}
\end{figure}

To clarify the possible anisotropy of the SC gap, we compare ARPES spectra measured at various $k_{\rm F}$ points. As visible in Fig. 3, the symmetrized ARPES spectra display two peaks irrespective of the $k$ location, demonstrating the absence of gap nodes. When we carefully look at the $k$ dependence of the $k_{\rm F}$ spectrum, we find a finite variation in the energy position of the quasiparticle peaks, suggesting the anisotropic character of the gap, which has not been well established in previous ARPES measurements on other Fe-based superconductors \cite{Hong, Zhou, Hasan, Nakayama1, Terashima, Nakayama2, Miao, Li}. As seen in Fig. 3(c), the peak position of the $\beta$ band moves toward higher $E_{\rm B}$ on going from the $\Gamma$-M direction ($\theta$ = 90$^{\circ}$) to the $\Gamma$-X direction ($\theta$ = 45$^{\circ}$). On the other hand, the peak energy of the $\gamma$ band shows a local maximum along the $\Gamma$-M direction ($\theta$ = 90$^{\circ}$) and it decreases while approaching the M-X direction ($\theta$ = 45$^{\circ}$) [Fig. 3(d)], suggesting that the anisotropy is rotated by 45$^{\circ}$ between the $\beta$ and the $\gamma$ FSs.  As for the $\delta$ FS, the energy position of the peak keeps a nearly constant value within the present experimental uncertainty [see Fig. 3(e)], suggesting a small anisotropy. To discuss more quantitatively the gap function of LiFeAs, we estimated the SC gap size $\Delta$ and plotted it as a function of the FS angle in Figs. 4(a) and 4(b). The results definitely confirm the multi-gap nodeless nature of the superconducting order parameter as well as the finite gap anisotropy on the $\beta$ and $\gamma$ FSs. Since the observed anisotropy is four-fold symmetric, we have performed a fitting by assuming $\Delta$($\theta$) = $\Delta_0$+$\Delta_1$cos[4($\theta$+$\phi$)] where $\Delta_1$ represents the magnitude of the gap anisotropy and $\phi$ reflects the phase shift of the gap function. As shown by solid curves in Figs. 4(a) and 4(b), the parameters of ($\Delta_0$ (meV), $\Delta_1$ (meV), $\phi$ (degree)) = (5.0$\pm$0.1, 0, 45), (2.6$\pm$0.1, 0.4$\pm$0.2, 45), (3.6$\pm$0.2, 0.6$\pm$0.2, 0) and (2.9$\pm$0.1, 0.07$\pm$0.1, 0) give a reasonable agreement with the experimental results for the $\alpha$, $\beta$, $\gamma$ and $\delta$ bands (corresponding magnitudes of the gap anisotropy are $\sim$0, 31$\pm$16, 33$\pm$13 and 5$\pm$7\%, respectively).

Now we discuss the implication of the present ARPES results in relation to the SC mechanism. Our results demonstrate: (i) anomalously strong-coupling behavior, (ii) FS-dependent nodeless SC gaps, and (iii) moderate gap anisotropy on some of the FS sheets. These findings strongly suggest an unconventional nature for the superconductivity in LiFeAs and the importance of anisotropic pairing interactions. A key question in understanding the SC mechanism is what kind of the gap symmetry is compatible with the experimental observation. Apparently, the experimental absence of gap nodes excludes the possibility of gap symmetries with vertical line nodes, such as the nodal $s$ wave, the $d$ wave, and the $p$ wave. A plausible pairing symmetry would be either the $s$ wave or the $s_{\pm}$ wave, which can be originated from the orbital \cite{Kontani, Yanagi} or the AF fluctuations \cite{Mazin, Kuroki, Seo}, respectively. One of the previous ARPES studies reported that the SC gap size is almost identical among the observed three holelike FSs on Ba$_{1-x}$K$_x$Fe$_2$As$_2$ and BaFe$_2$As$_{2-x}$P$_x$ \cite{Shimojima}, leading to an interpretation based on the $s$-wave pairing due to orbital fluctuations. However, the present ARPES result on LiFeAs showing a FS-dependent SC gap is obviously different from these results, but rather similar to other ARPES results that reported multiple SC gaps \cite{Hong, Zhou, Hasan, Nakayama1, Terashima, Miao}. Until now, no reasonable quantitative explanation based on the orbital fluctuation mechanism is available for the observed FS dependence of the SC gap.  To further evaluate the validity of the orbital-fluctuation model, it is highly desired to construct its theoretical gap function which can be directly compared to the present ARPES results.

It has been reported in previous ARPES studies \cite{Hong, Hasan, Nakayama1, Miao} that the FS-dependent SC gap is basically explained by the $s_{\pm}$-wave gap function $\Delta$($k$) = $\Delta_0$cos$k_x$cos$k_y$, derived from the local AF exchange coupling model \cite{Seo}.  This formula predicts a larger (smaller) gap on a smaller (larger) FS, qualitatively consistent with the present observation. In Fig. 4(c), we plot the experimentally determined gap values as a function of $\mid$cos$k_x$cos$k_y$$\mid$.  As one can clearly recognize, the FS dependence of the gap size basically follows the gap function with $\Delta_0$ = 4.7$\pm$0.4 meV, suggesting the importance of the AF interactions for the pairing.  The gap anisotropy along the $\beta$ FS also shows a good agreement with the $\Delta_0$cos$k_x$cos$k_y$ function.  

A remaining unresolved issue regarding the $s_{\pm}$-wave scenario is the anisotropy/isotropy along the $\gamma$ and $\delta$ FSs. While the appearance of gap maximum (minimum) along the $\Gamma$-M (M-X) direction on the $\gamma$ FS is qualitatively consistent with the $\mid$cos$k_x$cos$k_y$$\mid$ function, the experimentally observed anisotropy is substantially larger than that expected from the gap function. For the $\delta$ band, the experimental data show much smaller anisotropy as compared to the expectation. The origin of these finite deviations is still an open question. A hybridization between the two electron pockets may play some role. Indeed, Fig. 2(c) shows that the ellipses hybridizing to form the $\gamma$ and $\delta$ bands have a quite small eccentricity. Therefore, these bands must have mixed orbital characters over a wider range of FS angle, thus reinforcing elastic interband scattering between them, which may be detrimental to the SC pairing.  Accordingly, the observed deviation becomes most prominent around $\theta$ = 45$^{\circ}$ ($\mid$cos$k_x$cos$k_y$$\mid$ $\sim$ 0.8) where the $\gamma$ and $\delta$ bands are closest to each other ($i.e.$, the hybridization effect becomes the strongest).  Another aspect may be the mixture with another gap function.  For instance, by adding a small cos$k_x$+cos$k_y$ term, a gap anisotropy for the $\gamma$ FS might be produced, indicating that a more complex pairing interaction may be involved for the SC gap along this FS.

In conclusion, we reported our high-resolution ARPES results on LiFeAs ($T_{\rm c}$ = 18 K). We revealed that there are two holelike and two electronlike FSs at the $\Gamma$ and M points, respectively, where the SC gap shows a nodeless behavior in all the FSs. While the simple $s_{\pm}$-wave gap function of cos$k_x$cos$k_y$ can describe the overall FS dependence of the SC gap, a moderate gap anisotropy is observed along the outer hole and inner electron FSs, suggesting the complicity of pairing interactions in this material, possibly with the mixture with another pairing symmetry. Our observation of the detailed SC gap characters indicates the unconventional nature of the superconductivity in LiFeAs and puts a strong constraint on theoretical models proposed to explain the SC mechanism of the Fe-based superconductors.

We thank Xi Dai and Jiangping Hu for their valuable discussions and suggestions. We are also grateful to T. Kawahara for his help in the ARPES experiments. This work was supported by grants from JSPS, TRiP-JST, CREST-JST, MEXT of Japan, NSFC, the Chinese Academy of Sciences, NSF, Ministry of Science and Technology of China.

Note added. After completion of this work, we became aware of a related ARPES study on LiFeAs \cite{Borisenko2}, which reported similar gap anisotropy. Although that report concluded that the observed anisotropy is consistent with the orbital fluctuation scenario \cite{Yanagi}, our observation of a larger (smaller) gap opening along the $\Gamma$-M (M-X) direction of the inner electron $\gamma$ FS seems inconsistent with the theoretical prediction (Fig. 7 in ref. 7).

\end{document}